\documentclass[10pt]{article}
\usepackage{epsfig}
\usepackage{amsfonts}
\usepackage{rotating}
\usepackage{cite}

\newcommand{\beq}{\begin{equation}}
\newcommand{\eeq}{\end{equation}  }

\makeatletter
   \newcommand\figcaption{\def\@captype{figure}\caption}
   \newcommand\tabcaption{\def\@captype{table}\caption} 
\makeatother

\begin{document}

\hfill  ANL-HEP-CP-02-103

\begin{center}

\vspace{2.cm}
%%%%%%%%%%%%%%%%%%%%%%%%%%%%%%%%%%%%%%%%%%%%%%%%%%%%%%%%%%%%%%%
%                   1997                              %
%%%%%%%%%%%%%%%%%%%%%%%%%%%%%%%%%%%%%%%%%%%%%%%%%%%%%%%%%%%%%%%
{\Large\bf
Jet algorithms: a minireview} 
%%%%%%%%%%%%%%%%%%%%%%%%%%%%%%%%%%%%%%%%%%%%%%%%%%%%%%%%%%%%%%%

\vspace{1.cm}
S.V.Chekanov 

\vspace{1.cm}
HEP division, Argonne National Laboratory,
9700 S.Cass Avenue, \\
Argonne, IL 60439.
USA \\ Email: chekanov@mail.desy.de

\vspace{1.cm}
{\it Presented at the 14th Topical Conference on Hadron Collider
Physics (HCP2002), \\ 28 Sep - 4 Oct 2002, Karlsruhe, Germany}

\vspace{2.5cm}

\begin{abstract}
Many jet algorithms have been proposed in the past
to study the hadronic
final state in $e^{+}e^{-}$, $ep$ and $p\overline{p}$ collisions.
Here we review some of the most popular, mainly concentrating on the jet
algorithms used at HERA and TEVATRON.
\end{abstract}

\end{center}

\section{Introduction}

Jet algorithms are tools to reduce information on the hadronic final state
resulting from  high-energy collisions:
instead of analysing a large number of hadrons produced in an event,
one could focus on a relatively small number of jets. This helps
to concentrate on main features of the underlying physics, the
theory of quantum chromodynamics (QCD), as well as allows the reconstruction  
of heavy particles of the Standard Model.

Jets can be found without jet algorithms. A jet is simply a highly
collimated bunch of particles (calorimeter cells, tracks,
etc.) that can  easily be found after a visual analysis of 
events. However, to compare observables based on jet momenta with a theory,
one needs an objective and a unambiguous jet definition to be used
by experimentalists and theorists on an equal footing. 
In this article, a
few most popular definitions  of jet algorithms are reviewed.

\section{Requirements on jet algorithms}

The jet definitions should satisfy
the following requirements:

1) Predictions for jets
should be infrared and collider safe: i.e. a measured jet cross
section should not change if the original parton radiates a soft parton 
or if it splits into two collinear partons; 

2) The decision on which jet algorithm to use
has to be based on understanding
of the size of high-order QCD corrections. At fixed-order QCD, an
observable, $A$, can be expressed by a perturbation series in powers of the
strong coupling constant,
$A=A_{1}\> \alpha _{s}(\mu _{R}) + A_{2}\>\alpha _{s}^{2}(\mu _{R}) + 
B(\mu _{R})$,
where $B(\mu _{R})$ denotes missing high-order QCD terms, 
$\mu _{R}$ is the renormalisation scale used to deal with the
ultraviolet divergences ($A$ is independent of $\mu _{R}$).
To estimate the  contribution from unknown $B(\mu _{R})$, 
$\mu _{R}$ can be varied within some range.
If the renormalisation scale is set to the jet transverse energy, 
$\mu _{R}=E_{\bot }$, a typical variation presently adopted
to estimate the renormalisation scale uncertainty 
is $0.5E_{\bot }<\mu _{R}<2E_{\bot }$ \footnote{This range 
should be considered as the convention.}.  An optimal algorithm
has to have a small uncertainty associated with such variations.
This gives an indication  
that missing high-order QCD  contributions do not 
change significantly  the fixed-order theoretical predictions; 

3) Close correspondence with the original parton direction, since
the association of jets with hard partons is the basic assumption
when the theoretical predictions are compared to the data. This property
is essential when simple kinematical considerations are used to reconstruct
heavy particles from the invariant mass of two or more jets;

4) An optimal jet definition should have small hadronisation
corrections, as well as  small hadronisation uncertainties. 
At HERA, the transverse
energies of jets are relatively small, therefore, it is essential to understand
these two effects. The hadronisation correction factor, $C$, is evaluated
as the ratio $\sigma _{hadrons}^{MC}/\sigma _{partons}^{MC}$, where
$\sigma ^{MC}$ is the jet cross section obtained using  Monte Carlo
(MC) models generated for hadrons or partons. For an optimal jet algorithm, 
$C\sim 1$. Note that such correction factor used to multiply the     
fixed-order QCD cross sections is not fully justified for every observable: 
the parton level of MC models is fundamentally
non-perturbative because of the QCD cut-off used to deal with divergent
integrals, and the number of partons in MC models
significantly exceeds the multiplicity of partons for fixed-order
calculations. This correction was adopted only in case if:
a) a fixed-order QCD calculation and the corresponding parton-level MC
prediction well agree ($<5\%$
difference); b) the hadronisation correction
is not large ($<20\%$); c) the hadronisation
uncertainties are small ($<5\%$). The latter can be
found  by comparing  hadronisation corrections
evaluated using the Lund string fragmentation model with the
cluster fragmentation models, which are both implemented 
in MC simulations. 
Numerous results from HERA indicated
that  measured jet cross sections better agree  with the next-to-leading order
(NLO)  calculations
corrected using the MC hadronisation correction; 

5) Suppression of soft processes related to the beam remnants
(this will be discussed in more details below);

6) Small experimental uncertainties;

7) Simple to use in experimental analyses and in theoretical calculations.
Note that the same jet algorithm has to be uniquely defined
for experimental and theoretical calculation inputs, without any
additional modification.

First, we will discuss jet algorithms for $e^{+}e^{-}$ collisions
when there are no spectator jets
(see \cite{jetsinE+E-} for more details). 
In this respect, $e^{+}e^{-}$ jet algorithms are 
simpler than those for hadron collisions.    

\section{Clustering algorithms for $e^{+}e^{-}$}

The $e^{+}e^{-}$ collisions occur in the centre-of-mass frame, which
coincides with the laboratory frame. Thus, it is desirable
to find a two-particle distance measure which is
invariant under
the rotations. In this case, a good choice is the energy, $E_{i}$,
and the polar angle, $\theta _{i}$, of the $i$th particle.
The distance measure for two particles,
$d_{ij}^{2}$, can be defined as
$d_{ij}^{2}=2E_{i}E_{j}(1-\cos \theta _{ij})$,
where $\theta _{ij}$ is the opening angle between two particles.
More often, $y_{ij}=d_{ij}^2/E_{vis}^{2}$, with $E_{vis}$ being the visible
event energy, is used. This gives some cancelation
of errors between numerator and denominator. Note that when masses of
hadrons (partons) are set to zero, the variable $d_{ij}$ coincides with the
invariant mass of two particles. The reason for this choice is obvious:
particles tend to cluster closer in invariant mass in the
region of small momenta.

This distance measure was used by the JADE collaboration \cite{JADE}
to define jets in the following way: The algorithm starts
with the initial list of particles. The two particles are merged into
one, provided that their distance $y_{ij}$ is smaller than the desired
minimum separation, $y_{cut}$. This procedure is repeated until all
pairs of clusters have separations above $y_{cut}$.

It has soon been  realized that the fixed-order QCD corrections are
sizeable for the JADE algorithm. The explanation is following: soft gluons,
which are copiously radiated far apart, may have a small
distance measure ($E_iE_j\sim 0$). This leads
to a ``phantom'' jets which do not reflect the hardness of jets. It
is likely that this feature has a direct impact on the reconstruction
of heavy particles decaying into jets, since the use of JADE algorithm
for the reconstruction of $W$ bosons \cite{jetsinE+E-} and top quarks
\cite{chekanov} in $e^{+}e^{-}$ is less successful than for other
algorithms.

The solution was found by replacing the JADE distance measure
by the following construction: 
$d_{ij}^{2}=2\cdot \min (E_{i}^{2},E_{j}^{2})(1-\cos \theta _{ij})$,
which corresponds to the square of 
the transverse energy, $E_{\bot }^{2}$, of the
lower-energy particle with respect to a reference direction given
by the higher-energy parton, since for small angles 
$d_{ij}^{2}\sim 2\cdot \min (E_{i}^{2},E_{j}^{2})\cdot 
\sin \theta _{ij}^{2}=E_{\bot }^{2}$.
The jet-clustering based on this distance measure is called the Durham
or the $k_{\bot }$ algorithm \cite{KTalgor}. In this algorithm, the
soft gluons are combined first with the nearby high-order quark, thus 
the  $k_{\bot }$ algorithm 
avoids the problem of unnatural assignments of particles to jets. 

For $e^{+}e^{-}$, other algorithms, such as
LUCLUS, GENEVA, Angular-ordering Durham, CAMBRIDGE and DICLUS 
are also often  used (see \cite{jetsinE+E-} 
for details).

\section{Jet algorithms for $ep$ and $p\bar{p}$ collisions.}

\subsection{Differences between $e^{+}e^{-}$ and hadron collisions}

For more complicated colliding particles,
the initial-state system is not at rest and the laboratory frame is
less often used. The hadronic centre-of-mass frame and the Breit
frame (for $ep$ collisions in DIS regime) are the most natural choice.

There are a few reasons why the jet algorithms used in $e^{+}e^{-}$
cannot be applied directly to  collisions with more complicated initial
state:
a) In $e^{+}e^{-}$, the entire event arises from the collision, thus
one usually measures the \emph{exclusive} jet cross sections, i.e.  when  
all produced particles are grouped into jets and   
the cross sections  describe the
production of exactly $N$ number of jets and nothing else. In hadron-hadron
collisions, it is more convenient to analyse {\it inclusive} high $E_{\bot }$
cross sections, i.e. when some number of jets plus any number of unobserved
jets/particles are reconstructed. 
In this case, only a small fraction of the final-state hadrons
is associated with the large momentum transfer and hard scattering; 
b) The previous comment is easy to understand noting that  
the beam-remnant jet has huge energies, but it does
not undergone a hard scattering. Thus, the algorithm for hadron collisions 
should avoid
clustering particles with small transverse momenta with respect to
the beam direction, reducing contributions due to the ``underlying event'';  
c) Finally, in contrast to $e^+e^-$ events, where the rotation invariance
is important, for the hadron collisions one wants to emphasize the
invariance under the boost along the beam axis, as the 
partonic system is boosted along the direction of colliding hadrons. 
In this case, the separation
between particles can be defined in terms of the transverse energy,
$E_{\bot }$, azimuthal angle, $\phi $, and the pseudorapidity difference,
$\Delta \eta $ ($\eta =-\ln (\tan (\theta /2))$).

\subsection{The cone algorithm}

The cone algorithm has been used for a long time to define jets at
hadron colliders \cite{cone}. Every calorimeter cell with energy
above $E_{0}$ is considered as a seed cell (for the D0 choice, $E_{0}=1$
GeV). Then, a jet is defined by summing all cells within the cone 
$R_{cone}=\sqrt{(\eta _{i}-\eta _{seed})^{2}+(\phi _{i}-\phi _{seed})^{2}}$,
which is taken to be 0.7. 
The jet directions can be found as 
$\eta _{jet}=\sum _{i\in cone}E_{\bot i}\eta _{i}/E_{\bot {jet}}$,
$\phi _{jet}=\sum _{i\in cone}E_{\bot i}\phi _{i}/E_{\bot {jet}}$,
$E_{\bot {jet}}=\sum _{i\in cone}E_{\bot i}$. If the jet direction
does not coincide with the seed cell, the procedure is reiterated,
replacing the seed cell by the current jet direction, until a stable
jet configuration is obtained. After this, jets which are duplicated or
below some energy thresholds have to be thrown away. Since there is
no attempt to combine hadrons into the remnant jets, this  algorithm
is used to reconstruct inclusive jet cross sections. 
Some jets could be overlapping. 
To deal with this problem, the following procedure was adopted: any jet 
that has more than $50\%$ of its energy in common with a higher-energy
jet is merged with that jet (according to the D0 definition).
Any jet that has less than $50\%$ of
its energy in common with a higher-energy jet is split from that jet.
In case of the CDF and ZEUS algorithms, the energy merging/splitting
threshold is $75\%$. Note that after the merging/splitting procedure,
the size of the cone jets is not always equal to $R_{cone}=0.7$.

As it is clear from the above consideration, the cone algorithm is not
precisely defined, and there are many details which can affect the theoretical
results obtained using the cone-jet  definition. 
Thus, anyone calculating theoretical predictions must know the very precise
way of how this algorithm was implemented. In this respect,
\emph{clustering algorithms} to be discussed below do not suffer from
the ambiguities characteristic for the cone algorithm.

\subsection{The modified JADE algorithm}

This algorithm was one of the first algorithms used at HERA to reconstruct
jets and to determine the $\alpha_s$ values from the jet rates. 
Since the $ep$ hadronic final state is not as complicated as for hadron-hadron
collisions, one could slightly modify 
the JADE $e^+e^-$ algorithm by taking into 
account the new feature - the proton remnant
jet, but ignoring the requirement that the variables should be invariant
under the boost  \cite{JADEmod}:  
In order to cluster soft partons into the remnant jet,
a pseudo-particle which carries the missing longitudinal momentum
in the forward region was inserted. 
After the clustering, one ends up with $N+1$
jets (where "+1" denotes the proton-remnant jet). 
If no any experimental cuts on the jet kinematics are applied,
this algorithm can be used to measure the exclusive jet cross sections.

\subsection{The $k_{\bot }$ algorithm}

It is clear that the modified JADE algorithm has the same disadvantages
as the standard JADE algorithm for $e^{+}e^{-}$ annihilations: soft
gluons can be combined into phantom jets  even if the gluons are far apart.
Thus, the $k_{\bot }$ scheme should be used as a basis for the exclusive
jet definitions for hadron collisions. In contrast to the JADE algorithm,
however, it was proposed \cite{excKT} to use another method to deal
with the proton remnant jets in $ep$ collisions: 
one can define the distance from the proton direction
as $y_{k}=2\cdot (1-\cos \theta _{k})E_{k}^{2}/E_{\bot }^{2}$. 
Here, $E_{\bot}$ stands a hard scattering scale and $\theta _{k}$
is the angle of a particle with respect to the beam direction. Analogously,
$y_{kl}=2\cdot (1-\cos \theta _{kl})\cdot \min (E_{k}^{2},\, 
E_{l}^{2})/E_{\bot }^{2}$
can be defined for every particle pair. 
Then, the smallest value among $\{y_k , y_{kl}\}$ should be taken.
If $y_{kl}$ is the smallest and $y_{kl}<1$, two particles
are combined into a single cluster, $p_{kl}=p_{k}+p_{l}$. 
If $y_{k}$ is the smallest
and $y_{k}<1$, the particle is included into the beam jet. This
procedure is repeated until all clusters have $y_{k},\, y_{kl}>1$. The final
results are the remnant jet and some number of hard jets. This method was proposed
for the Breit frame of DIS.

This algorithm can also be used for $p\bar{p}$ collisions
if one adds  an additional distance measure for the 
second proton directions \cite{excKT}.

\subsection{The longitudinally invariant $k_{\bot }$ algorithm}

The two previous  clustering algorithms were designed as close as possible
to the clustering algorithms 
used in $e^{+}e^{-}$, i. e. they focus  on the exclusive
jet definitions. However,
it is possible to focus on the inclusive jet 
definition from the very beginning, by modifying the
jet clustering procedure.
In addition, one can redefine the distance measure
keeping similarity with the cone algorithm and using the longitudinally 
invariant variables for the distance measure.
Such an algorithm can be constructed in the following
way  \cite{incKT}: 
For each particle and particle pair, one should define $d_{i}=E_{\bot i}^{2}$ and 
$d_{ij}=R^{-2}\min (E_{\bot i}^{2},E_{\bot j}^{2})[(\eta _{i}-\eta _{j})^{2}+
(\phi _{i}-\phi _{j})^{2}]$, respectively 
($R$ is a free parameter). Then, one finds
the smallest of all the $d_{i}$ and $d_{ij}$. If $d_{ij}$ is the smallest, 
particles are merged into a new cluster. If the smallest is $d_{i}$, 
this particle should be removed from the list. This procedure continues
until there are no more particles/clusters, and as it proceeds, it
produces a list of jets with successively larger values of $d_i=E_{\bot i}^2$. 
After some cuts on the jet transverse energy,  only a few jets with 
high $E_{\bot}$ can be used for comparisons with theory. 
According to the perturbative calculations \cite{incKT},
if $R\simeq 1.35R_{cone} \simeq 1$,
the inclusive jet cross sections obtained 
with this algorithm are very close to those reconstructed using the 
cone algorithm.

Such a modification of the exclusive $k_{\bot}$ 
algorithm has also been proposed in
\cite{cat}, noting that the original $k_{\bot}$ algorithm is 
longitudinal invariant only in the small-angle limit.  However, it admits 
a longitudinal-boost-invariant extrapolation to large angles if 
the distance measure is
defined as   
$d_{ij}=\min (E_{\bot i}^{2},E_{\bot j}^{2})[(\eta _{i}-\eta _{j})^{2}+
(\phi _{i}-\phi _{j})^{2}]$.

\section{Differences between algorithms}

\subsection{Exclusive algorithms}

The major disadvantage of the JADE algorithm
is in its significant recombination scheme dependence; widely separated
soft partons can be clustered, even though these partons do not form
a pencil-like jet. As a consequence, this leads to large high-order
QCD corrections (for example, see \cite{Mirkes}).

\subsection{Inclusive jet algorithms}

The cone and the longitudinally invariant
inclusive $k_\bot$ algorithm 
allow the reconstruction of the inclusive jet cross sections.
Such cross sections are 
less informative than the exclusive ones reconstructed
with exclusive algorithms which force all particles into a given
number of jets.
Nevertheless, the inclusive jets are sufficient
for studies of hard QCD, since jets with high $E_{\bot}$ reflect
large momentum transfer. 
In addition, the high-$E_{\bot}$ jets     
have the hadronisation and  detector corrections 
significantly smaller than for jets with  low $E_\bot$.
The latter are less reliably reconstructed and might be attributed 
the hadron debris. Exclusive jet algorithms 
can also produce the inclusive cross sections 
by ignoring jets with low $E_{\bot}$. 

The  exclusive $k_{\bot }$ algorithm in the small-angle limit 
is identical to the longitudinally invariant $k_{\bot }$ 
algorithm: the difference appears  
for large angles when the longitudinally invariant algorithm is  
somewhat closely related to the cone algorithm. This simplifies the comparisons 
with the results obtained using  the cone algorithm. 
The major 
differences between the longitudinally invariant $k_{\bot }$ algorithm and the
cone algorithm are: 

1) The longitudinally invariant $k_{\bot }$ algorithm (as any other
algorithm based on the recombination procedure) never assigns a particle
to more than one jet, which is not the case for the cone algorithm;
for the latter,  
an arbitrary procedure is necessary to deal with this problem.

2) The distribution of transverse energy within jets is different.
The cone algorithm has well defined smooth boundaries
irrespective of the energy distribution of the 
hadronic activity inside the jet. 
This typically leads 
to more transverse energy near the cone edges than in case
of the longitudinally invariant algorithm. The latter can have rather 
complicated
boundaries depending on the energy flow within jet (see Fig.~1).
As a direct consequence, the resolution on the reconstruction of the
invariant masses of heavy particles from jets is better when the cluster
$k_{\bot}$ algorithm is used \cite{mike}.

3) The cone algorithm is not infrared safe \cite{Seym} 
at the next-to-next-leading-order QCD for $p\bar{p}$
process, when jets begin to develop internal structure. As a reflection
of this, the cone algorithm has large renormalisation scale uncertainties
already at next-to-leading order QCD calculations
for dijet cross sections in DIS (in
fact, the cross sections determined with the cone algorithm are negative
if the laboratory frame is used, indicating very large renormalisation
scale uncertainties).  A finite calorimeter cell size and
minimum energy $E_{0}$ used to define the seeds 
render finite  cone-jet cross sections. 
The seeds were  always considered 
by experimentalist as an insignificant
detail in the jet founding, since the seeds 
only help to find stable jet directions. 
However, the energies and the size of the 
seeds are very important for high-order QCD calculations, since 
the jet cross sections depend logarithmically on the energy threshold
above which calorimeter cells are considered as seed cells in the
overlap regions of two cone jets. 

In  contrast to the cone algorithm, the $k_\bot$ algorithm is
infrared and collinear safe to all orders of QCD calculations,
and it does not require the arbitrary splitting/merging procedure.

\section{Experimental situation}

The standard jet algorithm used by the D0 and CDF Collaborations is
based on the cone definition. Recently, for the first time, the D0  
used the longitudinally invariant $k_{\bot }$ algorithm to
measure inclusive jet cross sections  \cite{D0inc}. 
The obtained results indicated that the experimental
cross sections are rather different to those reconstructed with the
cone algorithm, although the theoretical predictions are very similar 
for both algorithms. This
might be attributed to different hadronisation corrections and/or 
to a contribution from spectator partons. The CDF and D0  plan
to use the $k_{\bot }$ algorithm in Run II.

At HERA, the cone algorithm (PUCELL,  
a rather similar to the CDF definition)
was  frequently  used in the past. At present, almost all jet physics
at HERA is based on the $k_{\bot }$ algorithm (see Fig.~2).
This algorithm significantly simplifies 
the data analysis, leads to small hadronisation corrections  $(<10-20\%)$
and hadronisation uncertainties  $(<3\%)$, as well as 
to small renormalisation scale dependence  $(<10-20\%)$. Experimental
uncertainties are usually smaller than the theoretical,
and are typically below $3-5\%$ 
for the jet transverse energies
$E_\bot \sim 15-30$ GeV. This ultimately allows high  precision
measurements. As an example, a recent determination of the strong coupling constant
from the inclusive jets at HERA \cite{ZEUSinc} has by a factor 
three less theoretical uncertainties than a similar measurement
based on the cone algorithm \cite{CDFinc}. 
Whether this can be attributed to
the use of the $k_{\bot}$ algorithm, or due to indisputable 
more complicated initial
state of colliding particles 
at TEVATRON is not yet clear and requires a careful examination.

In conclusion, it should be stressed that future developments 
of the jet algorithms
should mainly depend on understanding of multiple-gluon emissions
and high-order QCD contributions. 
The jet-algorithm definitions should not be 
motivated by efforts to minimize experimental-related effects, which
are nowadays significantly smaller than the renormalisation-scale dependence.
Note that future developments might be rather unexpected;
first steps  beyond the jet clustering algorithms have  already  been
undertaking \cite{tkachov},
focusing on instabilities of the jet clustering algorithms and indisputable
ambiguity of their definitions.

%%%%%%%%%%%%%%%%%%%%%%%%%%%%%%
\begin{minipage}{0.47\textwidth}
\begin{center}
        \includegraphics[height=6.0cm,angle=0]{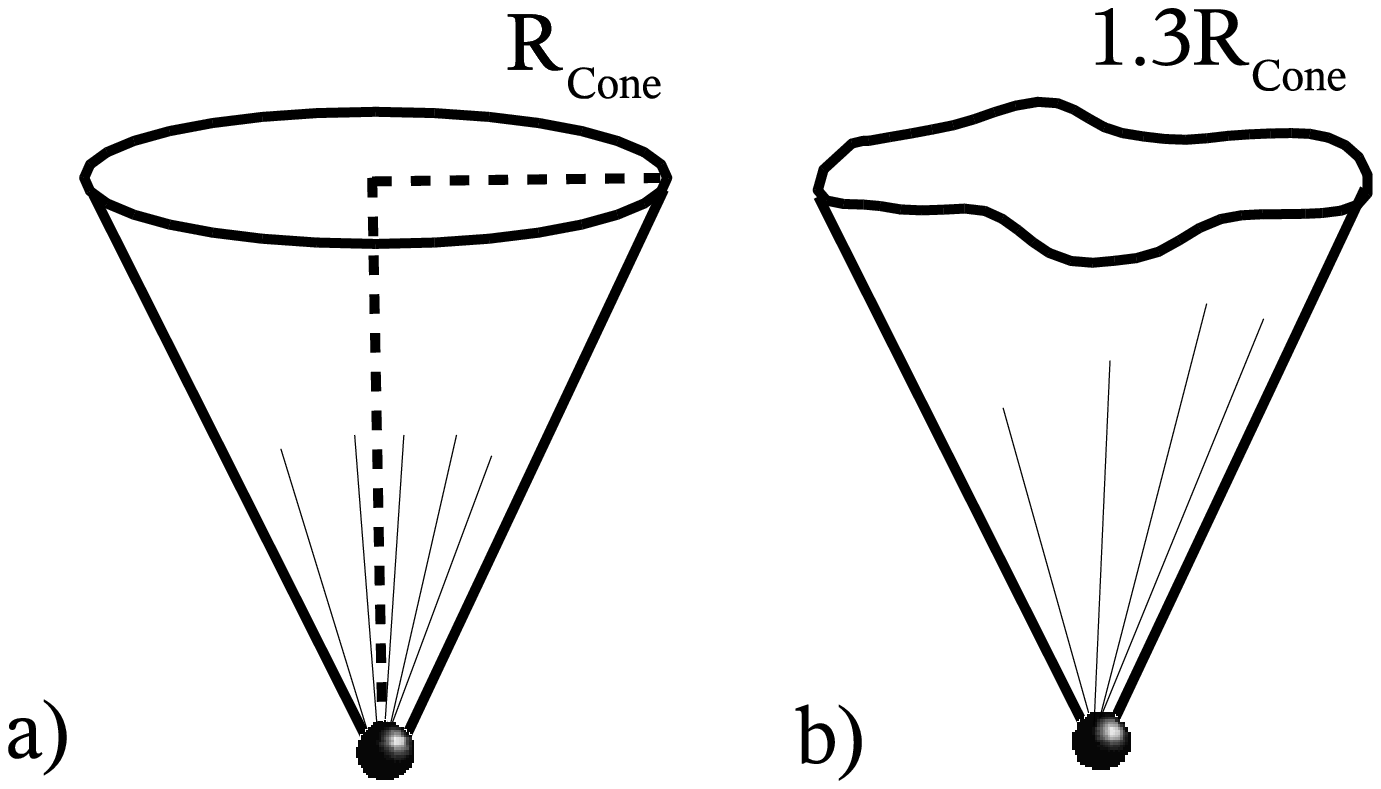}
\end{center}
\label{cap:Cone}
\figcaption{
Topologies of jet shape  for the cone (a) and for the longitudinally
invariant $k_\bot$ (b) algorithm.}
\end{minipage}
\hfill
\begin{minipage}{0.47\textwidth}
\begin{center}
        \includegraphics[height=6.4cm,angle=0]{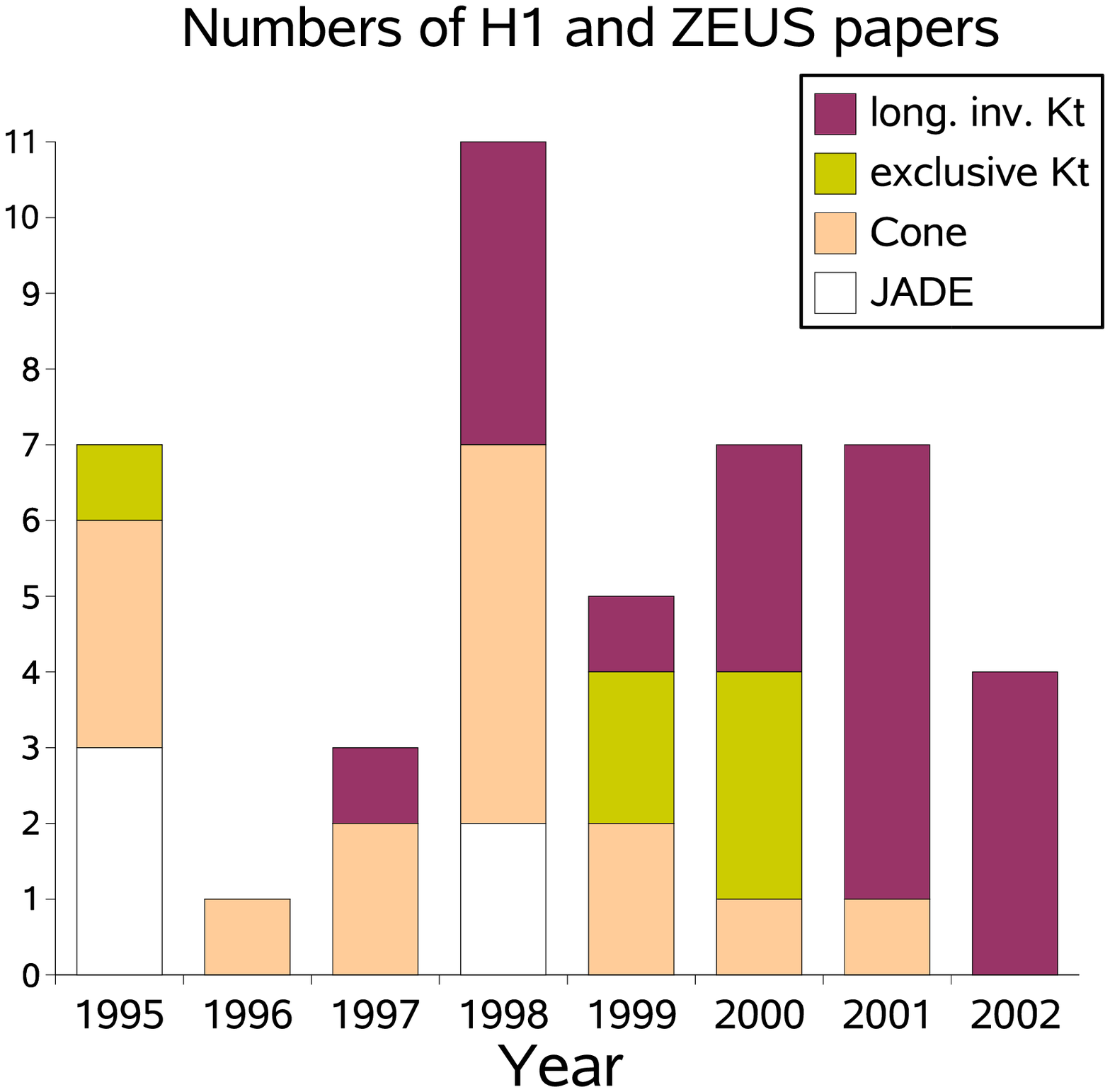}
\end{center}
\label{cap:The-numbers-of}
\figcaption{
The numbers of H1 and ZEUS published
papers based on jet algorithms.
Note that year of the articles does not always coincide  with  year
of publication.
}
\end{minipage}
%%%%%%%%%%%%%%%%%%%%%%%%%%%%%%%

\section*{Acknowledgments}
I thank J.Terr\'{o}n for discussions on this topic.

\end{document}